\documentstyle[12pt]{article}
\begin{document}
\begin{center}
{\Large {\bf QUASIELASTIC VERSUS INELASTIC AND DEEP INELASTIC LEPTON 
SCATTERING IN NUCLEI AT $x > 1$ }}
\end{center}

\vspace{0.5cm}

\begin{center}
{\Large {E. Marco and E. Oset. }}
\end{center}

\vspace{0.3cm}

{\small {\it
Departamento de F\'{\i}sica Te\'orica and IFIC, Centro Mixto Universidad
de Valencia - CSIC, 46100 Burjassot (Valencia) Spain.}}

\vspace{0.7cm}

\begin{abstract}
We have made a thorough investigation of the nuclear structure
function $W_{2A}$ in the region of $0.8 < x < 1.5$ and $Q^2 < 20 
$ GeV$^2$, separating the quasielastic and inelastic
plus deep inelastic contributions. The agreement with
present experimental data is good giving support to
the results for both channels. Predictions are made in yet unexplored
regions of $x$ and $Q^2$ to assert the weight of the quasielastic or
inelastic channels. We find that at $Q^2 < 4$ GeV$^2$ the structure 
function is dominated by the quasielastic contributions for $x < 1.5$, 
while for values of $Q^2 > 15$ GeV$^2$  and the range of $x$ studied
the inelastic channels are over one order of magnitude bigger than
the quasielastic one. 

The potential of the structure function at $x > 1$ as a source of
information on nuclear correlations is stressed once more.

\end{abstract}

\section{Introduction}
Deep inelastic scattering in nuclei has been a subject of intense study in
the past years, which was stimulated by the discovery of the EMC effect
\cite{1}. Different reviews on the subject have incorporated the advances
on the experimental and theoretical sides \cite{1A,2,3,4,5,6}. With ups and 
downs several conventional effects have remained as important ingredients 
explaining the basic features of the experiment: pionic effects for the 
enhancement of the ratio of the $F_2$ nuclear structure function to the one
of the deuteron \cite{4,7,8,9} around $x = 0.1$, binding effects for the 
depletion of that ratio around $x \simeq 0.6$ \cite{10,11,12} and Fermi motion
for the increase of the ratio for $x > 0.7 $ \cite{3,7,13}.

Relativistic effects were also shown to be relevant \cite{14,15} and the use
of spectral functions was advocated in refs.~\cite{16,17} as an 
important tool to accurately account for Fermi motion and binding effects.

In a recent paper \cite{18}, a theoretical framework was developed which 
accounts very accurately for all these effects. It uses a relativistic 
formalism from the beginning and follows a Feynman diagrammatic many body
scheme, where all the input from the nucleons or pions is incorporated via 
their respective propagators in the nuclear medium. The cloud from 
$\rho$-mesons was
also found relevant in \cite{18} and helped improve the agreement with
experiment. The scheme provided a good reproduction of the EMC data for
different nuclei outside the shadowing region which was not explored.

On the other hand, interest has been growing in the region of $x > 1$. This
region, inaccessible for free nucleons, clearly indicates nuclear effects
and the Fermi motion is certainly the first candidate \cite{12,13,19,20}. In
these latter works it was noted that a nonvanishing value of the nuclear 
structure functions at $x > 1 + k_F / M$ (with $k_F$ the Fermi momentum)
would indicate the presence of momentum components beyond the Fermi momentum
and this is actually the case experimentally.

The high momentum components of the nucleus are generated by the nuclear
correlations. For this reason the importance of these correlations
in generating the structure function at large values of $x$ has been
emphasized repeatedly \cite{1A,2,16a}, although the relevant information
on the spectral functions was not readily available at that time \cite{2}.
However, it is very dangerous to try to relate physical
quantities to the occupation number, $n (\vec{k})$, since the high momentum
components are strongly correlated with the energy of the nucleon and in
physical processes one has conservation of fourmomentum. The importance of
using the nucleon spectral function, $S_h (\omega, k)$, which provides the
probability of finding a nucleon with a certain energy $\omega$ and a 
momentum $k$, was pointed out in refs.~\cite{16,17,21,22,23}.

In ref.~\cite{24} a thorough study of the region of deep inelastic scattering
for \mbox{$x > 1$} was done for several nuclei. Several spectral functions were
used. One of them \cite{25} was calculated for infinite nuclear matter using
a microscopic Brueckner-Hartree-Fock scheme and the $NN$ interaction from 
realistic OBE potentials \cite{26}. A second one was calculated 
semiphenomenologically \cite{26a}, evaluating the nucleon selfenergy diagrams in a nuclear
medium but using input from $NN$ experimental cross sections and the 
polarization
of the $NN$ interaction in the medium to circumvent the use of the 
nucleon-nucleon potential and the ladder sums. In both cases the local
density approximation was used to evaluate the structure functions for finite
nuclei.

Another approach used a spectral function for finite nuclei \cite{27} for the
case of $^{16}$O. An interesting result from ref.~\cite{24} is that the
results with the local density approximation and those of the finite nuclei
differ by less than $3 \%$ in the region of the EMC effect and at
values of $x \simeq 1 $ or bigger the differences are also less than $8 \%$.
The differences between the results with the two spectral functions of nuclear
matter are of the same order of magnitude or smaller
than those quoted above. This gives us confidence
to rely upon the method in order to explore the $x > 1$ region where the 
information provided by the spectral function is essential. Indeed, it was
found in ref.~\cite{24} that usual approximations, like the use of the 
non interacting Fermi sea, or the use of the momentum distribution 
$n (\vec{k})$
provided by the same spectral function, but taking $\omega = E (\vec{k})$
(the free nucleon energy), or even an average energy $\omega (\vec{k})$
weighted by the spectral function, gave rise to results which differed by
two or three orders of magnitude from the accurate calculation and the
experiment.

The realistic calculation using the spectral functions was compared with the
only experiment at $x > 1$ in the deep inelastic region \cite{28} and the
results agree within $20-30 \%$. The data of ref.~\cite{28} explore only the
region of $x < 1.3$, and beyond 1.16 the data are only upper bounds. It was
also found in ref.~\cite{24} that the region of $x \approx 1.3 - 1.4 $ was
only sensitive to the high momentum and binding components provided by the
nuclear correlations and that the shell model part of the nuclear spectral
function gave a negligible contribution in that region.

It is thus clear that more data in the $x > 1$ region are necessary to further
study these interesting dynamical properties of nuclei beyond the simple shell
model picture \cite{29}. This is the same conclusion raised in ref.~\cite{6},
which states that high $Q^2$ data at high $x$ should be a top priority.

On the other hand, many more data at lower values of $Q^2$ exist, mostly
measured at SLAC \cite{30,31}, which could shed light on the same nuclear
issues, and others are planned at TJNAF \cite{32}. However, it was soon
noticed that these data have a large contamination of quasielastic contribution
where one nucleon of the nucleus is knocked out but there is no particle
production from this individual nucleon \cite{22,33A,33,35}. The presence of this
quasielastic background helped interpret the ``accidental" $\xi$ scaling 
observed in these reactions \cite{36}.

In ref.~\cite{33} three different regions are differentiated: the 
quasielastic,
the inelastic and the deep inelastic regions. The first one corresponds to
one nucleon removal, the second one to the excitation of low lying resonances
which decay by pion emission and the third one to deep inelastic where more
particles are produced. As quoted in ref.~\cite{33} these three regions are
never completely separated although one can find the dominance of some of
these channels in different regions. The investigations of ref.~\cite{33}
concluded that the region of $x > 1$ and $Q^2 < 3$ GeV$^2$ is dominated
by the quasielastic process and that $Q^2 > \; 20$ GeV$^2$ is a safe 
region where the deep inelastic process would dominate at $1 < x < 2$. 

The work of refs.~\cite{18,24} improves over the method used in 
refs.~\cite{16,17} in the use of a relativistic framework which 
makes unnecessary 
the flux factor proposed in ref.~\cite{37} and used in refs.~\cite{16,17}
in order to introduce relativistic effects in the nonrelativistic calculations.
It was shown in ref.~\cite{18} that the latter procedure led to different
numerical results than the consistent relativistic calculation. In addition,
the work of ref.~\cite{18} included the contribution from the modification
of the meson cloud in the medium, however, this feature is irrelevant in the
$x > 1$ region since the mesonic effects become negligible beyond $x > 0.6$.

With the advent of future experiments at intermediate energy machines like
TJNAF, or possible ones in proposed facilities like ELFE, it becomes 
important to make an exploration of that region in order to assess the 
dominance of the different processes at different $Q^2$ and $x$ and the
characteristics of the different processes. This is the aim of the present
paper which complements the findings of ref.~\cite{33}.

\section{Formalism for inelastic lepton scattering}

We follow closely the formalism of ref.~\cite{18} and write the inelastic
lepton nucleus cross section as

\begin{equation}
\frac{d^2 \sigma}{d \Omega d E'} = \frac{\alpha^2}{q^4} \;
\frac{k'}{k} \; L'_{\mu \nu} \; W'^{ \mu \nu}_A
\label{eq:1}
\end{equation}

\noindent
where $k, k'$ are the momenta of the incoming and outgoing lepton and $q$ 
the virtual photon momentum (see fig.~1). In eq.~(\ref{eq:1}), 
$L'_{\mu \nu}$ is the 
lepton tensor

\begin{equation}
L'_{\mu \nu} = 2 k_\mu k'_\nu + 2 k'_\mu k_\nu + q^2 g_{\mu \nu}
\label{eq:2}
\end{equation}

\noindent
and $W'^{ \mu \nu}_A$ is the hadronic tensor. We can use eq.~(39) of 
ref.~\cite{18} by means of which the nuclear hadronic tensor, in the 
local density
approximation, can be written in terms of the nucleon one as

$$
W'^{\mu \nu}_A = 4 \int d^3 r \int \frac{d^3 p}{(2 \pi)^3} \frac{M}{E (
\vec{p}\,)} \int_{-\infty}^\mu d p^0 S_h (p^0,p)
W'^{\mu  \nu}_N (p, q)
$$

\noindent
with

\begin{equation}
p \equiv (p^0, \vec{p}\,) ; \quad W'^{\mu \nu}_N = \frac{1}{2} 
(W'^{\mu \nu}_p + W'^{\mu \nu}_n)
\label{eq:3}
\end{equation}

By following the prescription of ref.~\cite{18}, $W'^{\mu \nu}_N (p, q)$,
which appears with nucleon variables off shell, is evaluated by taking the
matrix elements on shell $(E (\vec{p}), \vec{p})$, while the $\delta$
functions of conservation of fourmomentum are strictly kept with the off shell
variables. In eq.~(\ref{eq:3}) $S_h (p^0, p)$ is the spectral function 
for hole states of the correlated Fermi sea, normalized as

\begin{equation}
4 \int d^3 r   \int \frac{d^3 p}{(2 \pi)^3} \int^\mu_{- \infty}
S_h (\omega, p, k_F (\vec{r}\,) ) d \omega = A
\label{eq:4}
\end{equation}

\noindent
in symmetric nuclear matter, where $\mu$ is the chemical potential. 

Gauge invariance imposes the following structure of the hadronic tensor in 
terms of two invariant structure functions $W_1, W_2$,

\begin{equation}
W'^{\mu \nu} = \left( \frac{q^\mu q^\nu}{q^2} -
g^{\mu \nu} \right) W_1 + \left(p^\mu - \frac{p \cdot q}{q^2} q^\mu \right) 
 \left( p^\nu -
\frac{p \cdot q }{q^2} q^\nu \right) \frac{W_2}{M^2}
\label{eq:5}
\end{equation}

Furthermore, in the Bjorken limit $W_1$ and $W_2$ are related by the 
Callan-Gross relation and hence all information is given by one of 
the structure 
functions and $W_2$ is usually chosen for presentation of the data. In the
studies of inelastic scattering at lower values of $Q^2$ it is also customary
to show results for $W_2$,
hence we use eqs.~(3) and (5) to write $W_{2 A}$ in terms of $W_{2 N}$ by
eliminating the structure function $W_1$. This is easily accomplished by
using the expressions for $W'^{xx}$ and $W'^{zz}$ and taking  $q$ in the
$z$ direction, as usually done, and we get

$$
W_{2 A} = - \frac{q^2}{|\vec{q}|^2} \sum_{p, n} 2 \int d^3 r 
\int \frac{d^3 p}{(2 \pi)^3} \; \frac{M}{E (\vec{p})} \, 
\int^{\mu}_{- \infty} d \omega \; S_h (\omega, |\vec{p}|)
$$

\begin{equation}
\times \; [(p^x)^2 - \frac{q^2}{q^{0 2}} \; (p^z - \frac{p . q}{q^2} 
|\vec{q}|)^2 ]
\; \frac{W_{2 N} (p, q)}{M^2}
\label{eq:6}
\end{equation}

\noindent
where we have substituted a factor 2 of isospin in eq.~(\ref{eq:3}) by 
the explicit sum
over protons and neutrons (see also ref.~\cite{38C} for these Fermi
motion corrections).

In the Bjorken limit, when 
$- q^2 \rightarrow \infty, q^0 \rightarrow \infty$, we define the variables

\begin{equation}
x_N = \frac{- q^2}{2 p q} ; \; \nu_N = \frac{p \cdot q}{M} ; \; 
Q^2 = - q^2 ; \; x = \frac{- q^2}{2 M q^0} ; \; \nu = q^0
\label{eq:7}
\end{equation}

and the structure functions

\begin{equation}
\begin{array}{l}
\nu_N W_2 (x_N, Q^2) \equiv F_2 (x_N, Q^2)\\[2ex]
M W_1 (x_N, Q^2) \equiv F_1 (x_N, Q^2)
\end{array}
\label{eq:8}
\end{equation}

\noindent
where $F_1, F_2$ depend only on $x_N$ (there are smooth QCD corrections
depending on $Q^2$) and we  find

\begin{equation}
F_{2 A} (x) = \sum_{p,n} 2 \int d^3 r \int \frac{d^3 p}{(2 \pi)^3}
 \frac{M}{E (\vec{p})} \int_{- \infty}^{\mu} d \omega S_h
(\omega, |\vec{p}|\,) \frac{x}{x_N}  F_{2N} (x_N)
\label{eq:9}
\end{equation}

\noindent
which is the expression found in ref.~\cite{18} for the $F_{2A}$ structure
function in the Bjorken limit.

Eq.~(\ref{eq:6}) allows us to take into account the small corrections from
the terms with $p^x$ and $p^z$ in the bracket of the
formula, which become negligible in the Bjorken limit. We use this
equation to evaluate the inelastic plus deep inelastic part of the nuclear
structure function.

In the case of the quasielastic contribution to the structure 
function, $W_{2A}^Q$,
we evaluate it explicitly from the imaginary
part of the lepton selfenergy of the diagram of fig.~2. By following the
steps of section 3 of ref.~\cite{18} we can write:

\begin{equation}
\begin{array}{ll}
W'^{ \mu \nu}  = & \sum_{n,p} 2 \int d^3 r \int \frac{d^3 p}{(2 \pi)^3}
\frac{M}{E (\vec{p})} \int_{- \infty}^\mu  S_h (p^0, p) d p^0\\[2ex]
&  \frac{M}{E (\vec{p} + \vec{q})} \bar{\sum}_{s_i} \sum_{s_f}
\langle p + q | J^\mu | p \rangle \langle p +q | J^\nu | p \rangle^* 
\delta (q^0 + p^0 - E (\vec{p} + \vec{q}))
\end{array}
\label{eq:10}
\end{equation}

\noindent
where $J^\mu$ is the ordinary current for the coupling of a photon
to a nucleon \cite{39A}.

As advocated in ref.~\cite{18}, we follow here the philosophy of taking the 
matrix elements of the current between free spinors (dependent on 
threemomentum only) while keeping the argument of the $\delta$ function
with the variables strictly off shell as they are provided by
the nuclear spectral function.

In terms of the Sachs form factors $G_E , G_M$ we can write 
the matrix elements of eq.~(\ref{eq:10}) as:

\begin{equation}
\begin{array}{l}
\bar{\sum}_{s_i} \sum_{s_f} \langle p + q | J^\mu | p \rangle 
\langle p + q | J^\nu | p \rangle^* = \\[2ex]
= \left\{ \frac{p_x^2}{M^2} ( 1 - \frac{q^2}{4 M^2} )^{-1} [G_E^2 (q)
- \frac{q^2}{4 M^2} G_M^2 (q) ] - \frac{q^2}{4 M^2} G_M^2 (q) 
\right\} \quad \hbox{for} \,  \mu, \nu = x, x\\[2ex]
= \left\{ \frac{1}{4 M^2} (1 - \frac{q^2}{4 M^2})^{-1} [G_E^2 (q) - \frac{q^2}
{4M^2} G^2_M (q)] (2 E (p) + q^0)^2 - \frac{\vec{q}\,^2}{4 M^2}
G^2_M (q) \right\} \quad \hbox{for} \; \mu, \nu = 0,0
\end{array}
\label{eq:13}
\end{equation}

\noindent
$W_{2A}^Q$ is given by means of eq.~(\ref{eq:5}) for the nucleus at rest by

\begin{equation}
W_{2A}^Q = (\frac{q^2}{\vec{q}\,^2})^2 \, W'^{00}_{A,Q} -
\frac{q^2}{\vec{q}\,^2} W'^{xx}_{A,Q}
\label{eq:14}
\end{equation}

Hence, by means of eqs.~(10), (13), (14) we can evaluate the quasielastic
 structure function $W_{2A}^Q$.

For the $W_{2N}$ inelastic plus deep inelastic
structure function of the nucleon we take the
parametrization
of refs.~\cite{13,38} where there is a part corresponding to the excitation
of low lying resonances (usually called the inelastic part) and a smooth part 
for excitation in the continuum which would stand for the deep
inelastic part.

For the electric and magnetic form factors we take the dipole parametrization
of ref.~\cite{39}

\section{Results and discussion}

In figs.~3, 4, 5 we show the results for the structure function $W_{2A}$
($\nu W_{2A}$ in the figures) for $^{56}$Fe. The figures show the
strength of the structure function as a function of $x$, but
$Q^2$ is related to $x$ in the experiment since the data correspond to
the scattering of electrons with a fixed initial energy,
a fixed angle for the final electron and variable
energy for the final electron. In fig.~3 the initial energy is $E =
3.595 \, GeV$ and angle $\theta = 20^0$ \cite{30}. In this case at
$x = 1$ one has $Q^2 = 1.27 \, $ GeV$^2$ and $Q^2$ increases as $x$ 
increases.
In fig.~4 $E= 3.595 \, GeV$, $\theta = 39^0$ \cite{30}, and
$Q^2 = 3.1 \, $ GeV$^2$ at $x = 1$.
In fig.~5 $E = 5.12$ GeV, $\theta = 56.6^0$ \cite{31} and $Q^2 = 6.83$ 
GeV$^2$ at $x = 1$. The nucleon structure function has been taken from
ref.~\cite{38}.

In fig.~3 we can see the results for the quasielastic and inelastic
contributions. The quasielastic contribution is dominant in all
the range of the figure and peaks around $x = 1$ (for on shell
nucleons at rest we would have a $ \delta(x - 1)$ function). The
spread of the quasielastic contribution is due to Fermi motion
and binding.
The inelastic plus deep inelastic contribution is  small
(we will call it inelastic for simplicity in what follows) in all
the range of the figure compared to the quasielastic contribution. However,
at values of $x < 1$  the inelastic contribution becomes more relevant.
In fact one can see in the figure that at $x = 0.8$ the inelastic
contribution is about 70 $\%$ of the quasielastic one, and its inclusion
is necessary to obtain a good agreement with the data in that region.
For values of $x >1$ the strength of the structure function is
completely dominated by the quasielastic contribution. The agreement 
with experiment is rather good up to values of $x = 1.4$ and from
there on our results start diverging from the data.

The region of $x > 1.5$ in the figure can be filled up by excitation
of $2p 2h$ components, either by renormalizing the final nucleon 
propagator (we have taken a free propagator for the ejected nucleon) or
incorporating meson exchange currents (MEC) into the approach. It
is easy to see that the $2p 2h$ excitation is favoured in the $x > 1.5$ 
with respect to the $1p 1h$ excitation.
Indeed, by splitting the fourmomentum $q^0, \vec{q}$  into two equal
halves with $q^0/2, \vec{q}/2$ for each $ph$ excitation as an
average, we see that this latter combination is kinematically much
more suited than the excitation of a $ph$ component with $q^0, \vec{q}$.

The contribution of these $2 p 2h$ components has been the
subject of intense investigation at lower electron energies 
\cite{40,41,42,43}. It is a genuine many body contribution which
is not accounted for by the inelastic contribution and hence is additional
to the quasielastic and inelastic channels studied here (the final nucleon
renormalization actually redistributes the strength and is not an
additive  channel, contrary to the case of the MEC).

We will not discuss further this subject since our purpose here is to
make a comparison of the quasielastic with the inelastic
contribution as a function of $Q^2$ and $x$ to establish the regions
where the inelastic contributions dominate in the cross sections.

In fig.~4 we find similar features to those discussed above except
that we see that the contribution of the inelastic channels is relatively
more important than in the former case. Indeed, at $x = 0.95$ the
quasielastic and inelastic contributions are about the same. Once again we
can see that the inelastic contribution is essential to describe
quantitatively the data below $x = 1$ and at $x < 0.9$ the inelastic
contribution dominates the structure function.

In fig.~5  we show the results for the recent experimental
data of ref.~\cite{31} taken at higher electron energies and $Q^2$.
At this value of $Q^2$ we see that the quasielastic and inelastic
contributions are similar around $x = 1.2$ but at higher and lower
values of $x$ the inelastic contribution is larger, particularly
at $x \leq 1$ where it becomes clearly dominant.

We can see from all these figures that the agreement of our results
with the data in the regions of dominance of the inelastic or
quasielastic channels is rather good and we then extrapolate the
results to make predictions for values of $Q^2$ still unexplored.

One interesting finding of ref.~\cite{24} is the sensitivity to
the nuclear spectral function $S_h (\omega, p)$ of the 
results for the structure functions $F_{2A}$ at $x > 1$ and $Q^2$ large, 
and how some common
approximations made in the study of the EMC effect badly fail at 
values of $x = 1.2 - 1.5$. Here we want to carry out a similar test for
the quasielastic channel and for the inelastic one at
these lower values of $Q^2$.

In fig.~6 we show the results of the quasielastic channel calculated
with the spectral function and with two approximations:

1) Non interacting Fermi sea, for which the spectral function can be
written as \cite{24}

\begin{equation}
S_h^{FS} (\omega, p, \rho) = n_{FS} (\vec{p}) \delta (\omega -
E (\vec{p}) - \Sigma)
\label{eq:17}
\end{equation}

\noindent
where $n_{FS}$ is the occupation number of a non interacting Fermi
sea of local density $\rho (r)$.

The magnitude $\Sigma$ in eq.~(\ref{eq:17}) is a selfenergy which is 
chosen such as to
provide the exact experimental binding energy of each 
particular nucleus \cite{24}.

2) Momentum distribution. In this case we
use eq.~(\ref{eq:17}), with a momentum distribution
$n_I (\vec{p}\;)$ which is the actual momentum distribution
in the nucleus given by

\begin{equation}
n_I (\vec{p}\,) = \int_{- \infty}^\mu S_h (\omega, p) d \omega
\label{eq:20}
\end{equation}

This approximation misses the correlations of the momenta with the
energy and we shall call it uncorrelated momentum distribution.

In fig.~6 we see that the use of the non interacting Fermi
distribution leads to similar results (slightly bigger) of the structure
function around the peak of the distribution (at $x = 1$) than
those using the spectral function. However, as $x$ increases, the results
with the non interacting Fermi sea fall faster than the other ones.

The spectral function accounts for larger momentum components than
the non interacting Fermi sea and this is the reason for the extended
contributions at $ x > 1$. We also show in the figure the results
obtained by using the uncorrelated momentum distribution of 
eq.~(\ref{eq:20}). We can see
that this approximation fails to provide the main features of the
quasielastic peak, which are well given by the non interacting Fermi sea
calculation or the one using the spectral function.
We also observe that at $x > 1.6$ the approximation gives rise to an 
important contribution which is not substantiated by the accurate
calculation with the spectral function. This is due to the contribution
of large momentum components which, however, are uncorrelated with
similarly large binding energies, as the spectral function would
give \cite{24}.

In fig.~7 we see similar results but for the inelastic contribution
corresponding to the data of fig.~5. In this case
the roles are reversed. We can see that the non interacting Fermi
sea gives results smaller than those with the spectral
function at $x >1$ and the uncorrelated momentum distribution gives rise
to results much bigger than those obtained with the spectral function
at $x > 1$. The discrepancies are rather large, such that none of these
approximations can be advocated as a fair substitute of the use of the full
spectral function. Particularly, the results obtained with the 
momentum distribution are rather bad in both  channels, the quasielastic
and the inelastic.

 Another way of
expressing these ideas is to say that the inelastic contribution at
$x > 1$ is rather sensitive to the nuclear spectral function and carries
information on this magnitude, particularly, as discussed in 
ref.~\cite{18}, about the ``background'' contribution which is tied to
the dynamical features of the nucleus.

Now we proceed to extrapolate the results at higher $Q^2$.
In fig.~8 we show the results for the quasielastic and inelastic contributions
at $x = 1$ as a function of $Q^2$. We use three different  
parametrizations of the nucleon structure function, one of them which we 
have used so far \cite{38}, and which is indicated for relatively low
$Q^2$ and other two, \cite{mrs} (MRS) and \cite{cteq} (CTEQ),
which are more suited for large $Q^2$ values
in the Bjorken scaling region. 
The figure is significative because we see that the results obtained
for the inelastic contribution are very sensitive to the parametrization
used for the nucleon structure function. At $Q^2 = 20$ GeV$^2$ the 
results obtained  with different nucleon structure functions differ
by about a factor 5. However, at values of $Q^2 \simeq 1$ GeV$^2$ the
differences are about a factor 50. A discussion of the assumptions
made in these structure functions and their range of validity is hence
in order. The structure function from CTEQ which we use is meant
to be used at large $Q^2$ in the Bjorken scaling region, since we
do not implement $Q^2$
corrections important at small $Q^2$. 
The structure function of MRS is meant to work at high values
of $Q^2$ and also at low values and hence, in principle, should cover the range
of $Q^2$ in fig.~8. However, data at small $Q^2$
and large $x$ are not included in the fit of MRS. Indeed,
the largest $x$ considered is $x = 0.85$ and the data for this 
value of $x$ go down to $Q^2 = 10$ GeV$^2$ only.

On the other hand, the parametrization of ref.~\cite{38} is precisely meant 
for low values of $Q^2$ and it contains the contribution from excitation
of resonances (the so called inelastic part) plus
a background of deep inelastic. This parametrization is done with precission,
observing also the thresholds. For instance $F_{2N} (x)$ vanishes before
$x = 1$ as it should be: Indeed, we have

\begin{equation}
s = Q^2 (\frac{1}{x} - 1) + M^2 \, > (M + m_\pi)^2
\label{eq:21}
\end{equation}

\noindent
since we need to create at least a pion  in the inelastic contribution.
Hence, in the limit $Q^2 \rightarrow \infty$, we have
$ 0 < x < 1$ and the limit
of $x = 1$ can be approached as much as one wishes.
However, for $Q^2$ of the order of a few GeV$^2$ there is a cut off at values
of $x < 1$, for example at $Q^2 = 1 \, $ GeV$^2$ , we have $x < 0.78$.
The parametrization of \cite{38} has a structure function vanishing 
in the forbidden region,
 while the one of MRS provides a smooth function of $x$ which
reaches the limit $x = 1$ for all values of $Q^2$.

Since the nuclear structure function $F_{2A}$ calculated at $x > 1$ picks
up its contribution  from $F_{2N}$
for values of $x_N$ close to 1, it is then
clear that the structure function of MRS should not be used for 
such purposes for low values of $Q^2$. Instead, the nucleon
structure function of \cite{38} should be used in this case. But the
latter one should not be used at large values of $Q^2$.

A compromise region can be $Q^2 \simeq 6-7$ GeV$^2$ below which the
\cite{38} results should be taken and beyond which the MRS should be
used. The  problems with the MRS and CTEQ
parametrizations to describe the large $x$
region can be further exposed in figs.~9, 10, 11.

In fig.~9 we show the results for $F_{2A}$ at  $x = 0.8$ with the three 
parametrizations. We see now a better agreement between the MRS and
\cite{38} parametrizations in the region of 3 GeV$^2 \leq
Q^2 \leq 8$ GeV$^2$. Below $Q^2 = 3$ GeV$^2$ we see again that the
MRS gives a sharp increase with respect to \cite{39} which again
must be attributed to the artificial large $x$ dependence
of $F_{2N}$. At values of $Q^2 > 15$ GeV$^2$ the CTEQ and MRS
parametrizations start converging.

The discrepancies between the results of $F_{2A}$, calculated with the 
different nucleon structure functions at $x = 1$ in fig.~8, become even
worse at higher values of $x$. In fig.~10 we show the results
at $x = 1.3$. Discarding the results obtained with the CTEQ parametrization
in that region, we still see that there are large discrepancies between
the MRS and \cite{38} results with a gap in the region of 
$Q^2 \simeq$ 6--7 GeV$^2$ which should in principle be the dividing region
for the two results. It is difficult to draw firm conclusions in the
region of 4 GeV$^2 < Q^2 < 15$ GeV$^2$, but we can safely claim that at 
values of $Q^2 < 4$ GeV$^2$ the quasielastic contribution dominates
the nuclear structure function, while at $Q^2 > 15$ GeV$^2$ the
inelastic part dominates.

The situation becomes even worse at $x = 1.5$ (not shown here). 
Once again it is difficult
to make up one's mind in the intermediate region of $Q^2$, but
in any case one could conclude rather safely the dominance of the inelastic
channel at $Q^2 > 20$ GeV$^2$ and of the quasielastic one at 
$Q^2 < 3$ GeV$^2$. These results agree with those obtained in 
ref.~\cite{33}.

It is clear that going to values of $x > 1.5$ would make the results even more 
sensitive to the parametrization of $F_{2N}$, thus increasing more
the intrinsic uncertainties of the nuclear structure function
and not allowing the extraction of useful nuclear information.

There is, however, good news if one goes to high values of $Q^2$ in the
Bjorken region. Indeed, as we show in fig.~11, at $Q^2 \simeq 85$
GeV$^2$ the nuclear structure function $F_{2A}$ is all due to the inelastic
contributions and the results obtained with different parametrizations
agree remarkably among themselves, which gives us great 
confidence in the results. We have carried out the calculations using three 
parametrizations: the MRS, the one of CTEQ and the one of Duke 
and Owens \cite{45}. The differences are of the order of 10$\%$ or
less. We also show in the figure the experimental results of 
ref.~\cite{28} The results agree with the data within 30$\%$ in the range
of $0.8 < x < 1.3$. As discussed in ref.~\cite{24} the results of
$F_{2A}$ at $x >1$ are essentially due to the ``background'' part 
of the nuclear spectral function, with a negligible contribution of
the bound states of the shell model. The structure function hence
contains a very rich information on ``nuclear correlations''
in the generalized sense of nuclear structure beyond the mean field
approximation.

Coming back to our discussion about the accuracy of the different
nucleon structure functions to provide $F_{2A}$ at $x >1$ we show in 
fig.~12 the results obtained for $Q^2 = 1.27\, $ GeV$^2$ (at $x = 1$) with
the MRS structure function. We can see that the addition of the
inelastic contribution to the quasielastic one leads to poor agreement
with the data. This confirms our previous statements that at
low $Q^2$ the MRS should give an overestimate of $F_{2A}$ because it does not
respect the thesholds in $x$.

On the other hand, in fig.~13 we show the results calculated with the
MRS structure function at $Q^2 = 6.83$ GeV$^2$ (at $x = 1$).
The agreement with the data in this case is of the same quality as
with the Stein structure function \cite{38}, slightly better at high 
values of $x$ and 
slightly  worse at smaller $x$. This figure substantiates our
statement that $Q^2 \simeq 6-7$ GeV$^2$ is the dividing line for the
validity of the two parametrizations and values of $x$
around 1 or below. At larger values of $x$ we saw, however, that in
that region there are discrepancies between the results with the
two structure functions.

Another message of our results is that if one wants to
extract useful nuclear information at values of $x > 1.3$ from the 
intermediate 
$Q^2$ region (3 GeV$^2  < Q^2 < 15$ GeV$^2$), a reliable nucleon
structure function in the region of $x$ close to 1 is absolutely necessary.
This call is important in view that this range is bound to be covered in
future experiments at TJNAF.

It is interesting to compare these results with those found in 
ref.~\cite{33}. The results found here at $x = 1$ are very similar to those
found in ref.~\cite{33}, qualitatively and quantitatively.  At
values of $ x > 1$ the trend of our results agree qualitatively with
those of ref.~\cite{35}. However, our discussion and the use of 
different nucleon structure functions served to show that at
present there are large uncertainties in the intermediate  $Q^2$ region
tied to the behaviour of the structure functions at $x$ close to
1, which prevent us from drawing firm conclusions in that region.

\section{Conclusions}

We have made a thorough study of the quasielastic and inelastic
contributions to the nuclear structure functions in the region
of $x \geq 1$ and for values of $Q^2$ ranging from 1- 20 GeV$^2$. This
is the region where both mechanisms compete and we have shown the
regions in the $x, Q^2$ variables where one or the other of the
mechanisms dominates. As discussed in the former section, in some
cases we reconfirm results obtained by other authors before,
but there are a few novelties in the present work. With respect
to ref.~\cite{24}, where only deep inelastic scattering at
high $Q^2$ was studied, we study here the region of small and 
intermediate $Q^2$ and include the quasielastic contribution which was 
negligible at high $Q^2$. This has allowed us to compare our theoretical 
predictions with many existing data, which had not been done before, and 
also has allowed us to make some prospects of what can be learned and
what should then be measured in present facilities like TJNAF.

With respect to the similar work of refs.~\cite{33,35} our results
include relativistic corrections. Furthermore, our study has shown that 
there is no precise information available on the nucleon structure functions
around $x=1$ in order to make reliable predictions in the region of
$6$ GeV$^2<Q^2<20$ GeV$^2$. We also give arguments which set clear limits
to the region of validity of present parametrizations of the nucleon
structure functions. On the other hand, even if we have different numerical
results at $x>1$ than in \cite{33,35}, the ``safe'' conclusions obtained here 
(taking into account the uncertainties) about the regions of
dominance of the quasielastic or deep inelastic contributions are
the same. We describe now with some more details the main findings
of the present work.

On general grounds we can say that at
$x = 1$ the quasielastic contribution dominates below $Q^2 < 3$
GeV$^2$  while for $Q^2 > 13$ GeV$^2$ the inelastic contribution
is already  an order of magnitude bigger than the quasielastic one.

For values of $x > 1.3$ we found that in the region of 3 GeV$^2
< Q^2 < 15$ GeV$^2$ the results for the inelastic contribution
obtained with different nucleon structure functions were very
different. This does not allow one to draw strong conclusions from
the data in that region.

We also observed that at values of $Q^2 \simeq 85$ GeV$^2$ the results
obtained for $F_{2A} $ at $x > 1$ with different nucleon structure functions
 were very stable. This fact, together with the sensitivity of these
results to the region of large momenta and binding energies of the nuclear
spectral function, makes the measurement of $F_{2A}$ in that region
an excellent tool to learn about dynamical aspects of the nucleus,
beyond its approximate shell model structure.

Even with the uncertainties about the inelastic contribution
in the intermediate $Q^2$ region, we can make some relatively safe 
statements by claiming that for $x < 1.5$ the inelastic contribution
is dominant for $Q^2 > 20$ GeV$^2$ while the quasielastic contribution
is dominant for $Q^2 < 4$ GeV$^2$. These results confirm the findings
of ref.~\cite{33}.

Our determination of the quasielastic contribution is rather precise
in the region of $Q^2 < 20$ GeV$^2$. This means that if experiments
are done in that region, our results can be used to separate the 
inelastic contribution to the structure function from the data. These
results would be useful to unravel the discrepancies between 
different models for $F_{2N}$ at $x$ close to 1. Certainly, precise
measurements of $F_{2N}$ in that region would also be needed
simultaneously if one wishes to asses the validity of the many
body methods used here. Taking into account that this range of
$Q^2$ is the one likely to be investigated at TJNAF, both types of 
experiments are called for.

With respect to the reliability of the method used, there are two regions
where our predictions for the inelastic contribution are safe: the region
of $Q^2 < 5$ GeV$^2$ where the parametrization for $F_{2N}$ of \cite{38}
used here is rather reliable, and the one in the deep inelastic
region at $Q^2 > 80$ GeV$^2$ where the results obtained are rather stable,
quite independent of the parametrization used.

At large $Q^2$ the agreement of our results with the only existing data is 
fair. At low values of $Q^2$ the agreement of our results with
the data is rather good and by changing the range of $Q^2$ and $x$
we change the strength of the inelastic and quasielastic contributions,
obtaining in all cases a good agreement with the data (at $x < 1.5 \,,
Q^2 < 5 \, $ GeV$^2$). This gives us some confidence in our method to
describe both the quasielastic and inelastic contributions in those
regions.

Another interesting information which one could get experimentally is
the separation between the quasielastic contribution and the inelastic
one. Here, however, we must give an anticipated warning about how
the comparison should be made. In the evaluation of the quasielastic 
contribution we did not look at final state interaction (FSI). This is fine
is one wishes to obtain the contribution from this channel to the $F_{2A}$ 
structure function, which sums the contributions of all
possible final states. In practice, the strength of the quasielastic
channel evaluated by us would be redistributed in other channels due
to final state interaction of the emerging nucleon with the nucleus,  and
hence some events which were quasielastic originally will become inelastic
ones in the nucleus due to FSI. This is in complete analogy with the
process of photon absortion in nuclei, which can proceed 
either through direct photon  absorption by pairs (or trios) of
nucleons or indirectly via pion production and pion reabsorption in
the nucleus \cite{46}. In  this case, events which were pion production
originally become nuclear absorption events, but the important thing
is that the inclusive cross section can be calculated by looking only at the
first step processes. The FSI does not change the cross sections obtained
in that way, it simply redistributes the strength in other channels.
Certainly, one of the possible measurements would be one nucleon
emission and the rest of the nucleus in its ground state. This is not a
trivial measurement since it requires a good energy resolution 
to guarantee that the final nucleus is not further excited. Provided
this experiment is done, the comparison of our results with this data
would require the use of distorted waves for the  emitted nucleon,
instead of the implicit plane wave calculation which we have done. On
the other hand, if what one wishes is  to separate the inelastic
contribution to be compared with ordinary evaluations of the
deep inelastic
nuclear structure function, what one must subtract from the data is not this
experimental one (and only one) nucleon removal contribution, 
but the quasielastic scattering
contribution evaluated by us, corresponding to the first step 
quasielastic scattering on one nucleon.

Summarizing our thoughts on future perspectives: the measurements at $x > 1$ 
in the high $Q^2$ region, $Q^2 \simeq 80$ GeV$^2$ should be encouraged.
They offer a clean measure of the deep inelastic contribution and
they provide direct information on the dynamical aspects of the nucleus,
loosely speaking, about nuclear correlations. Certainly other 
theoretical calculations should be most welcome.

The region of low $Q^2$, $Q^2 < 5$ GeV$^2$ and $x \leq 1$ seems to
be rather well under control theoretically but there the quasielastic
contribution is large. The
quasielastic contribution at low values of $Q^2$ is only sensitive to 
nuclear correlations for $x > 1.5$ where other contributions would be 
important, hence it can not be considered an important
 source of information on the dynamical aspects
of the nucleus. Conversely, the inelastic and deep inelastic contributions,
even at intermediate values of $Q^2$ carry this relevant information.
The existence of experimental facilities where this region of $Q^2$
will be explored can rend this information easily accessible. However,
for the purpose of capitalizing the information contained in these data,
the measurement of nuclear structure functions at $x  >1$ in that
region should be accompanied by precise measurements and accurate
parametrizations of the nucleon structure function at $x$ close to 1.

\vspace{2cm}

{\bf Acknowledgements:}

We would like to acknowledge useful discussions with W. Weise.
This work has been partially supported by DGICYT contract number AEN 93-1719.
One of us, E.~M., wishes to acknowledge a fellowship from the Ministerio 
de Educaci\'on y Ciencia.

\newpage

\begin{center}
\underline{Figures captions}.
\end{center}

\vspace{0.5cm}

$\bullet$ Figure 1: Feynman diagram for the inelastic lepton-nucleon
 scattering.

$\bullet$ Figure 2: Lepton selfenergy diagram associated with the quasielastic
 lepton-nucleon scattering.

$\bullet$ Figure 3: Results for $\nu W_{2A}/A$ for $^{56}$Fe 
 ($Q^2 = 1.27$ GeV$^2$ when $x=1$). Dot-dashed line: inelastic contribution
 using the nucleon structure function of \cite{38}; dashed line:
 quasielastic contribution; solid line: whole calculation including the
 inelastic and quasielastic contributions. Experimental points
 from ref.~\cite{30}.

$\bullet$ Figure 4: Same as in fig.~3 but in this case 
 $Q^2 = 3.1$ GeV$^2$ when $x=1$.

$\bullet$ Figure 5: Same as in fig.~3 but in this case
 $Q^2 = 6.83$ GeV$^2$ when $x=1$ and the experimental points are
 from ref.~\cite{31}.

$\bullet$ Figure 6: Results of the quasielastic channel using different
 approximations ($Q^2 = 1.27$ GeV$^2$ when $x=1$). Solid line: spectral
 function; dashed line: non interacting Fermi sea, 
 eq.~(\ref{eq:17}); dot-dashed line: uncorrelated momentum 
 distribution, eq.~(\ref{eq:20}). Experimental points
 from ref.~\cite{30}.

$\bullet$ Figure 7: Results of the inelastic channel using different
 approximations ($Q^2 = 6.83$ GeV$^2$ when $x=1$). Solid lines: spectral
 function; dashed line: non interacting Fermi sea, eq.~(\ref{eq:17}); 
 dot-dashed line: uncorrelated momentum distribution, eq.~(\ref{eq:20}). 
 Experimental points from ref.~\cite{31}.

$\bullet$ Figure 8: Results for the quasielastic and inelastic contributions
 for $^{56}$Fe at $x=1$ as a function of $Q^2$. Solid line: quasielastic
 contribution; long-dashed line: inelastic contribution using the CTEQ
 parametrization; short-dashed line: inelastic contribution using the
 MRS parametrization; dot-dashed line: inelastic contribution using the
 parametrization that includes the low lying resonances and a continuum 
 part \cite{38}. 

$\bullet$ Figure 9: Same as in fig.~8 but for $x=0.8$.

$\bullet$ Figure 10: Same as in fig.~8 but for $x=1.3$.

$\bullet$ Figure 11: Results for the structure function of $^{12}$C at
 $Q^2=85$ GeV$^2$ using different parametrizations for the nucleon
 structure function. Solid line: MRS \cite{mrs}; long-dashed line:
 CTEQ \cite{cteq}; short-dashed line: Duke and Owens \cite{45}. 
 Experimental points from ref.~\cite{28}.

$\bullet$ Figure 12: Results for $\nu W_{2A}/A$ for $^{56}$Fe 
 ($Q^2 = 1.27$ GeV$^2$ when $x=1$). Dot-dashed line: inelastic contribution
 using the MRS parametrization of the nucleon structure function; dashed line:
 quasielastic contribution; solid line:whole calculation including the
 inelastic and quasielastic contributions. Experimental points
 from ref.~\cite{30}.

$\bullet$ Figure 13: Same as in fig.~12 but in this case
 $Q^2 = 6.83$ GeV$^2$ when $x=1$ and the experimental points are
 from ref.~\cite{31}.

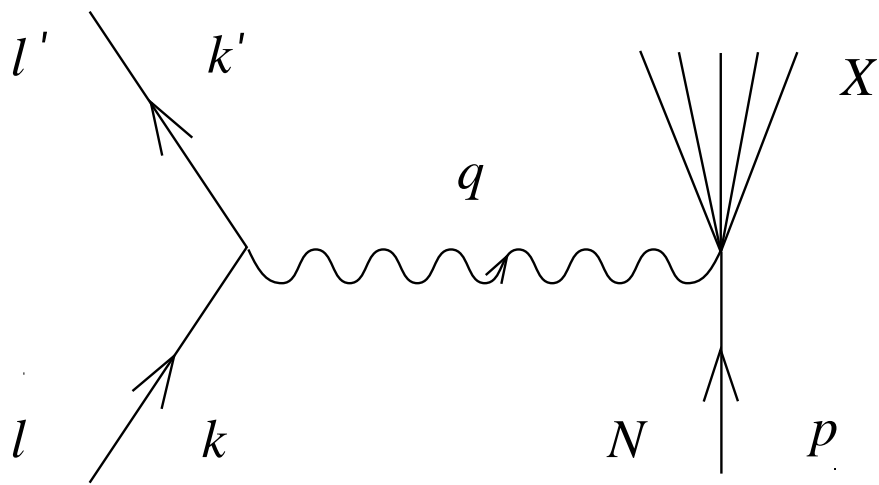
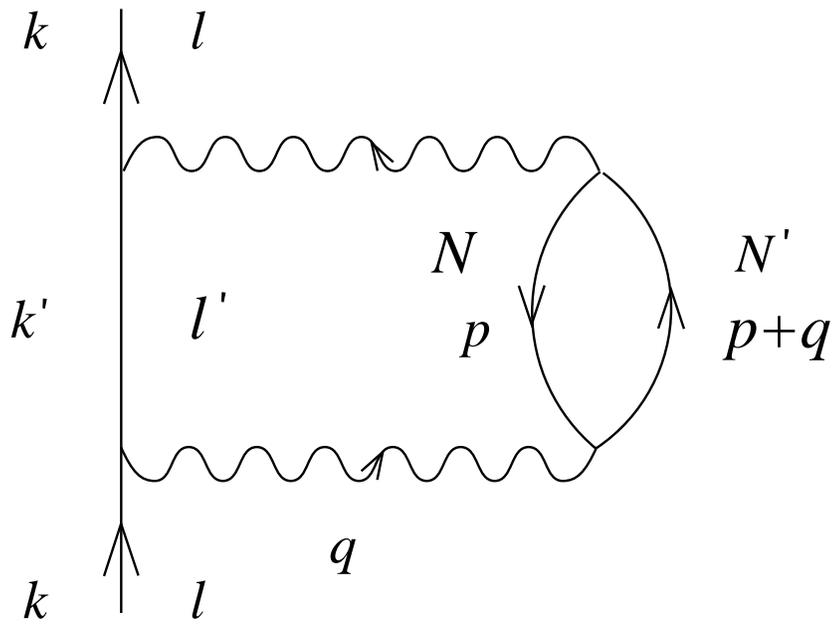
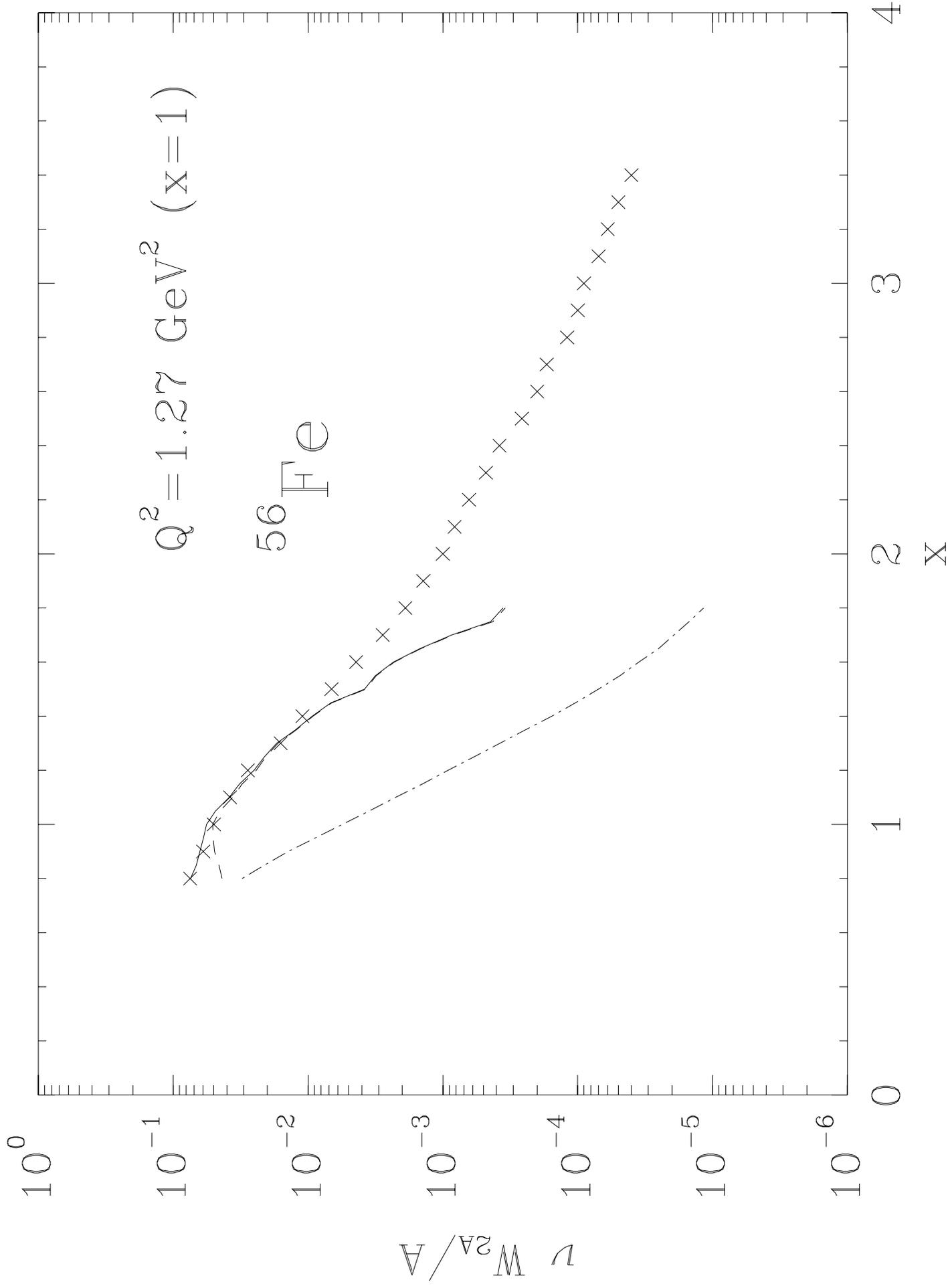
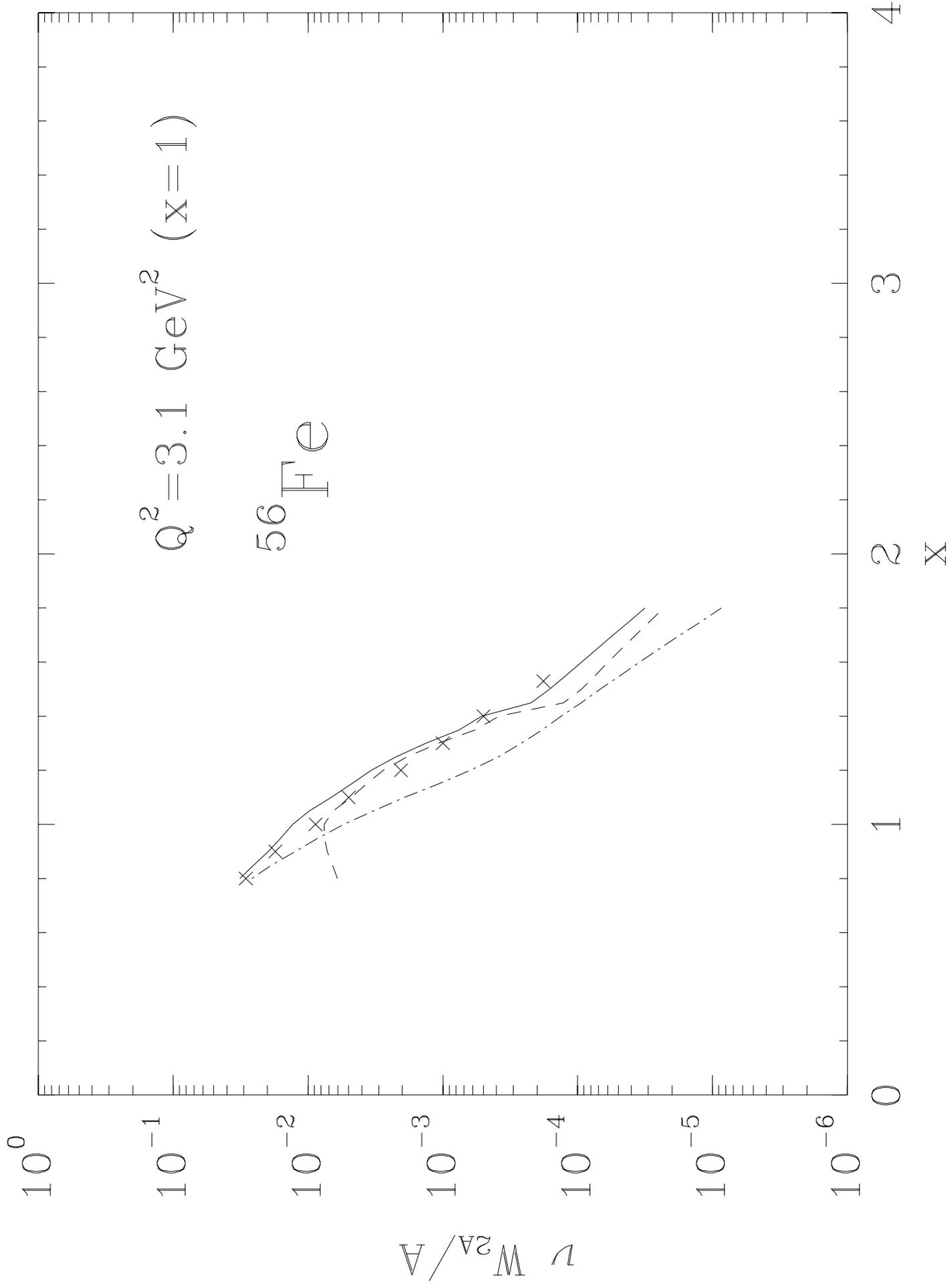
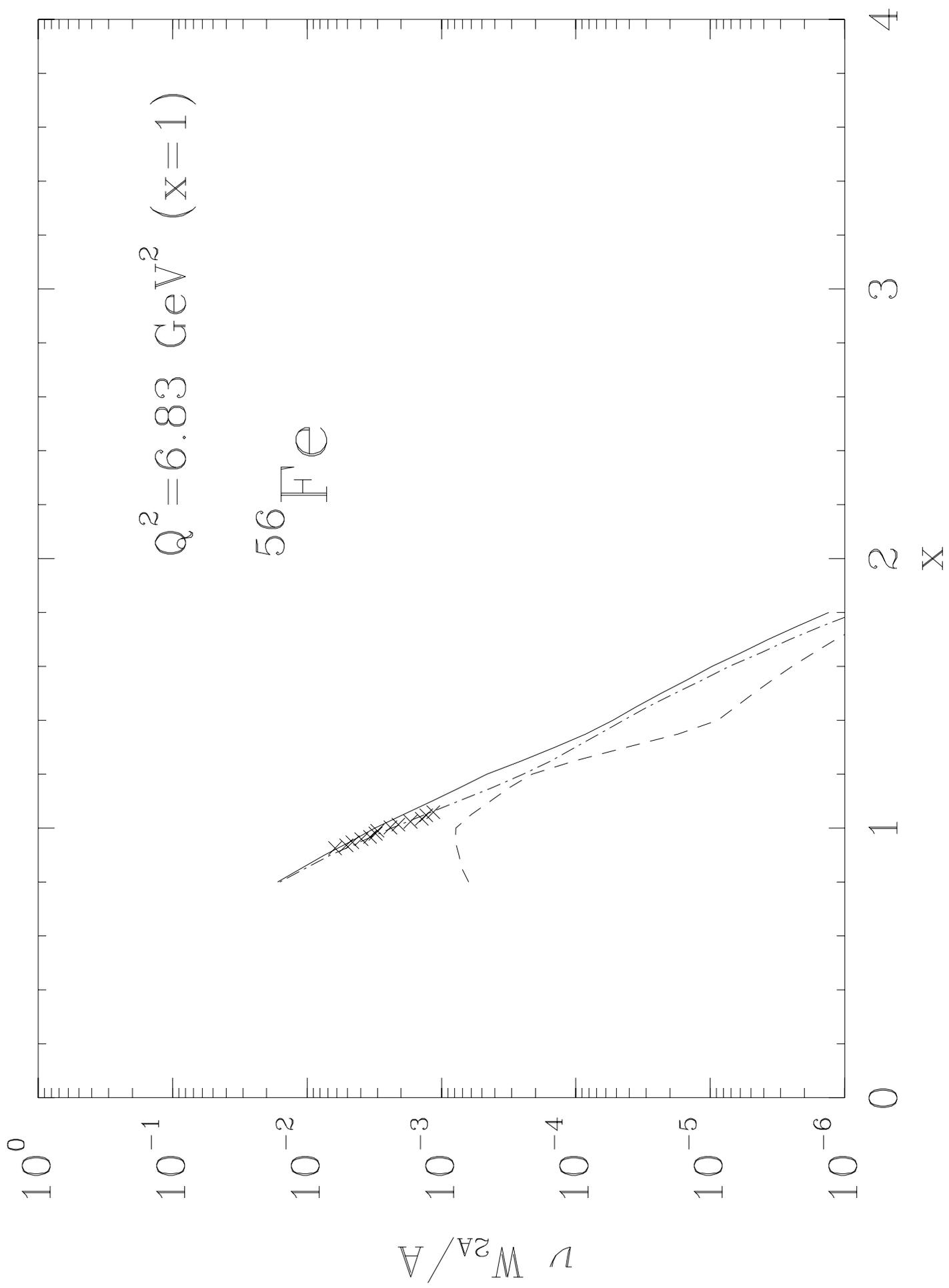
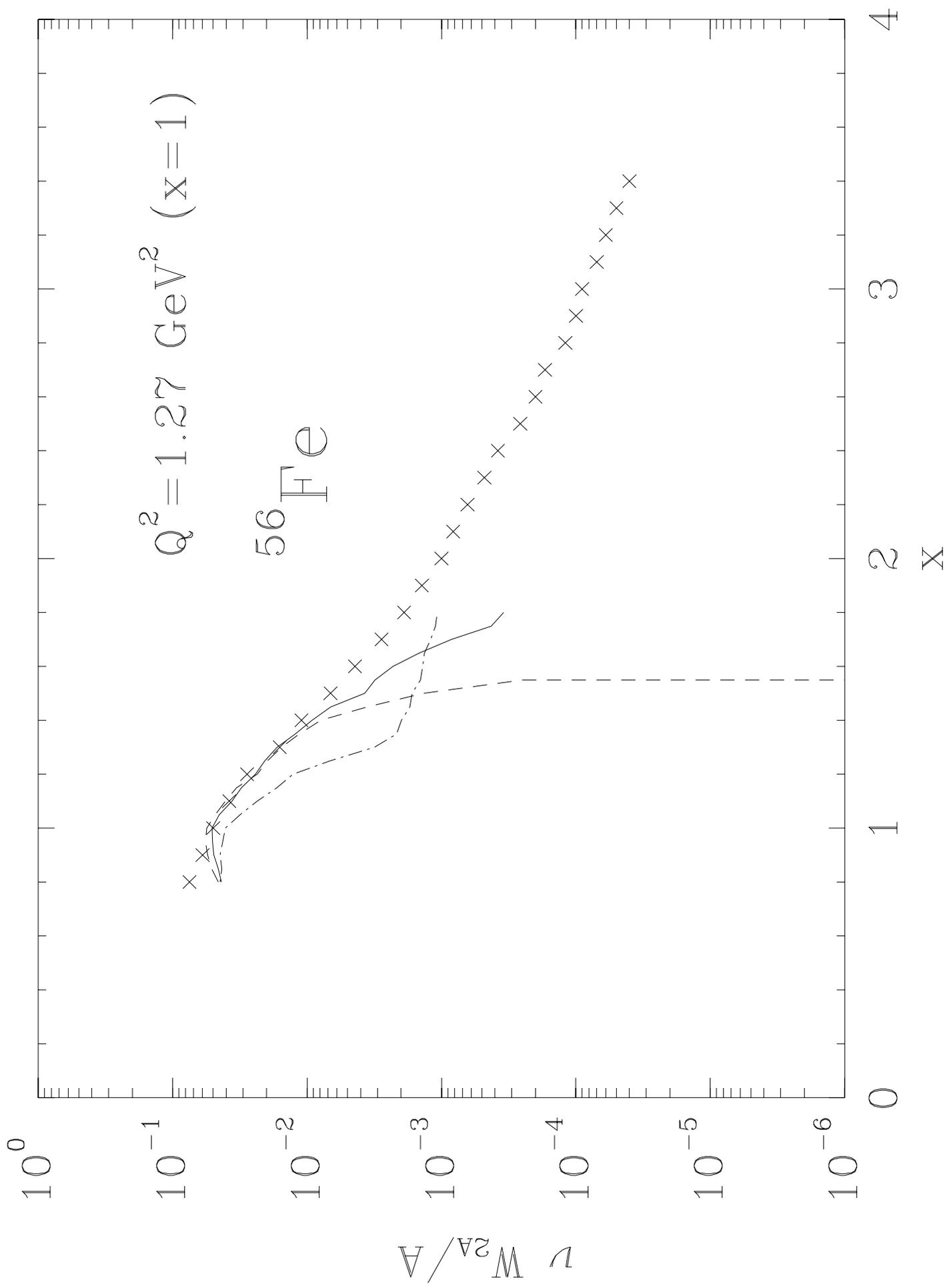
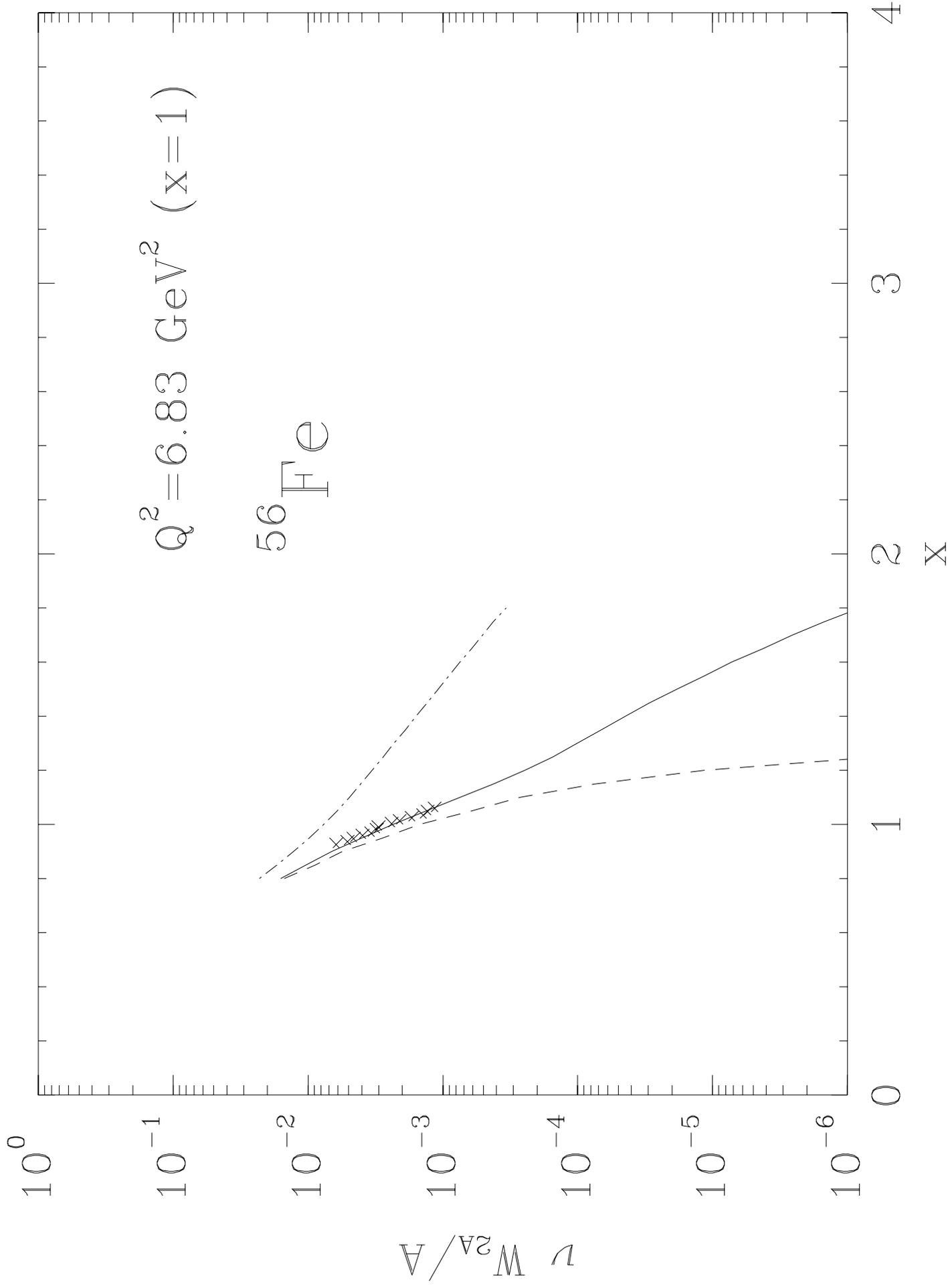
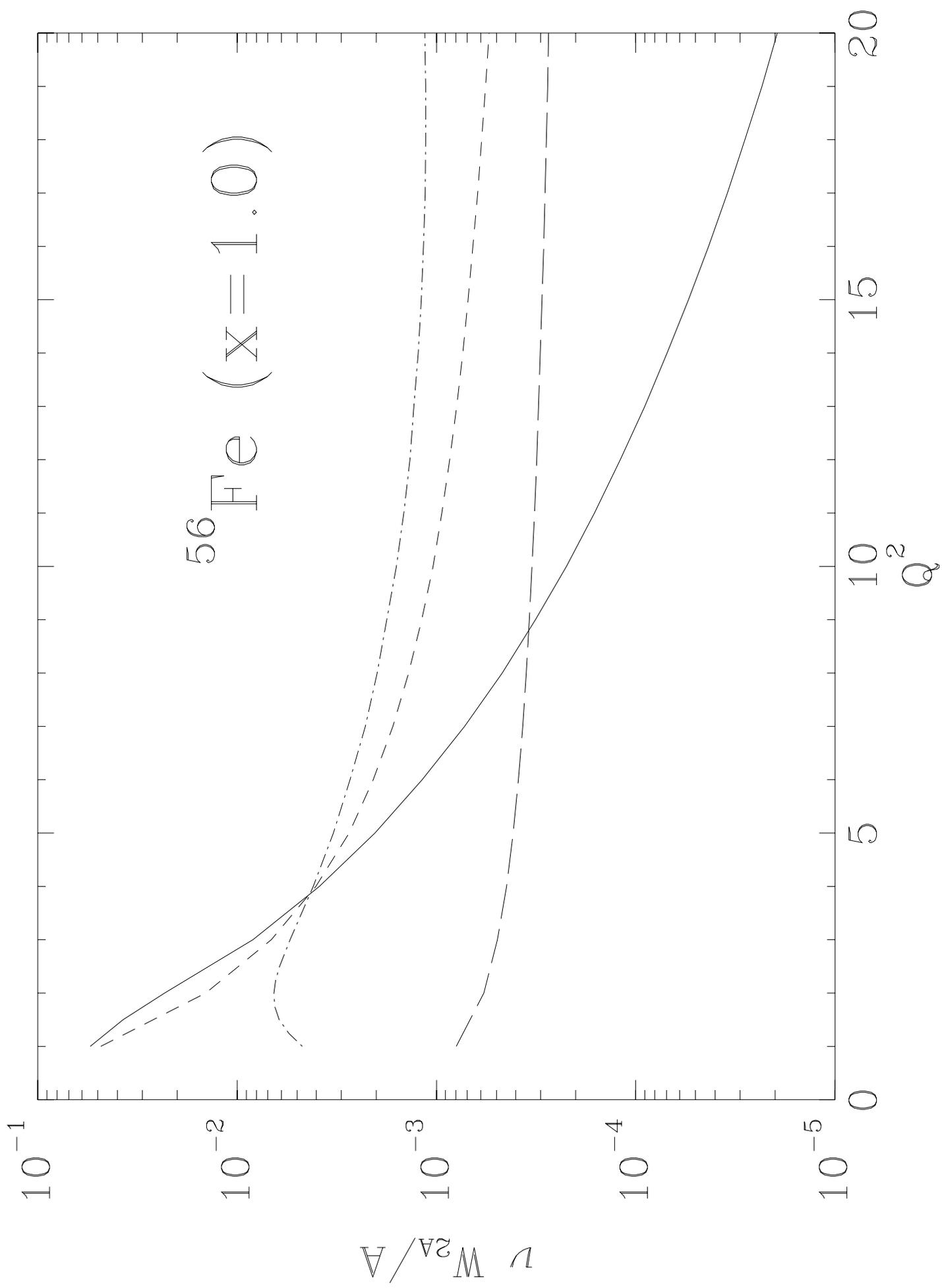
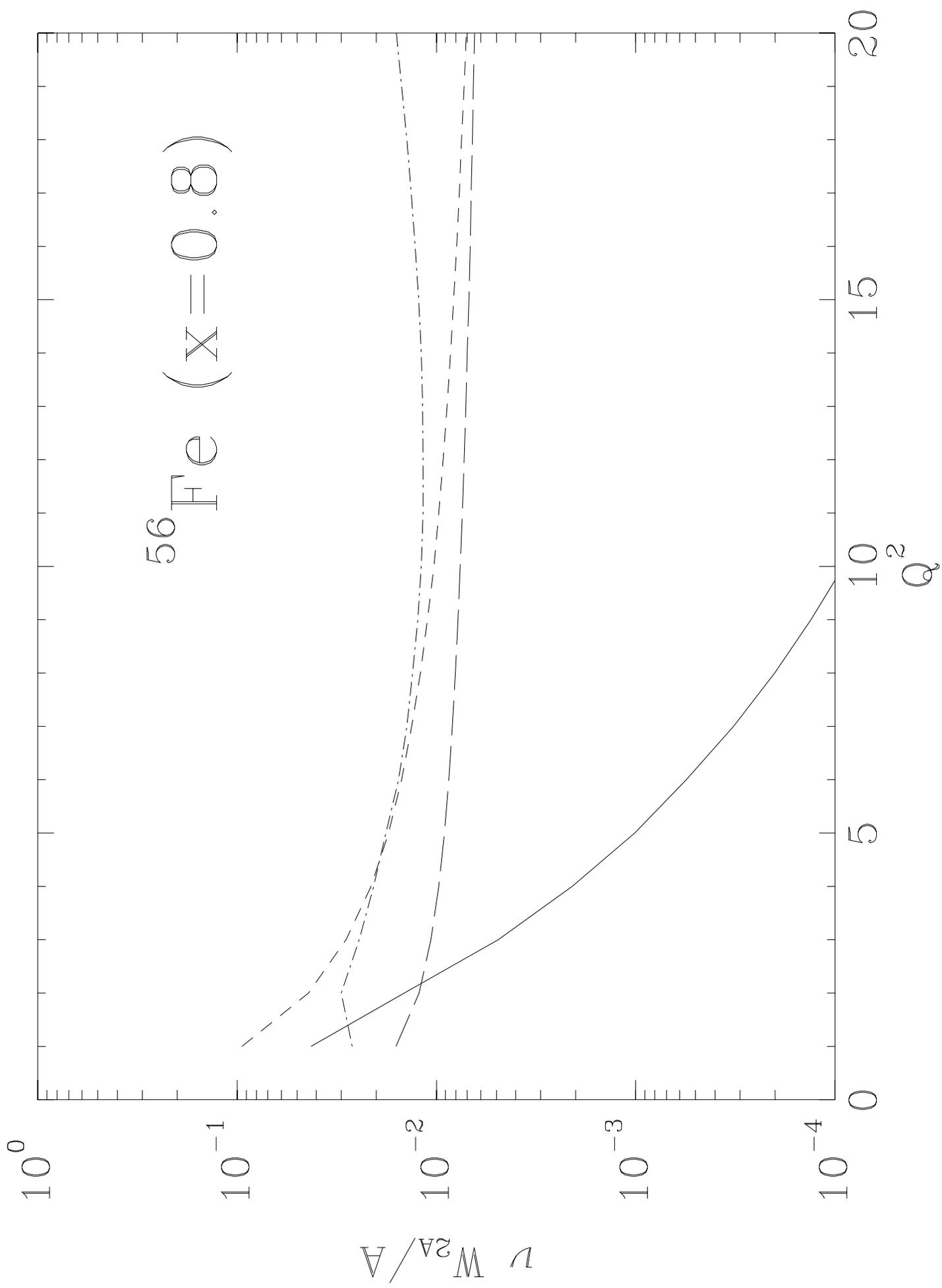
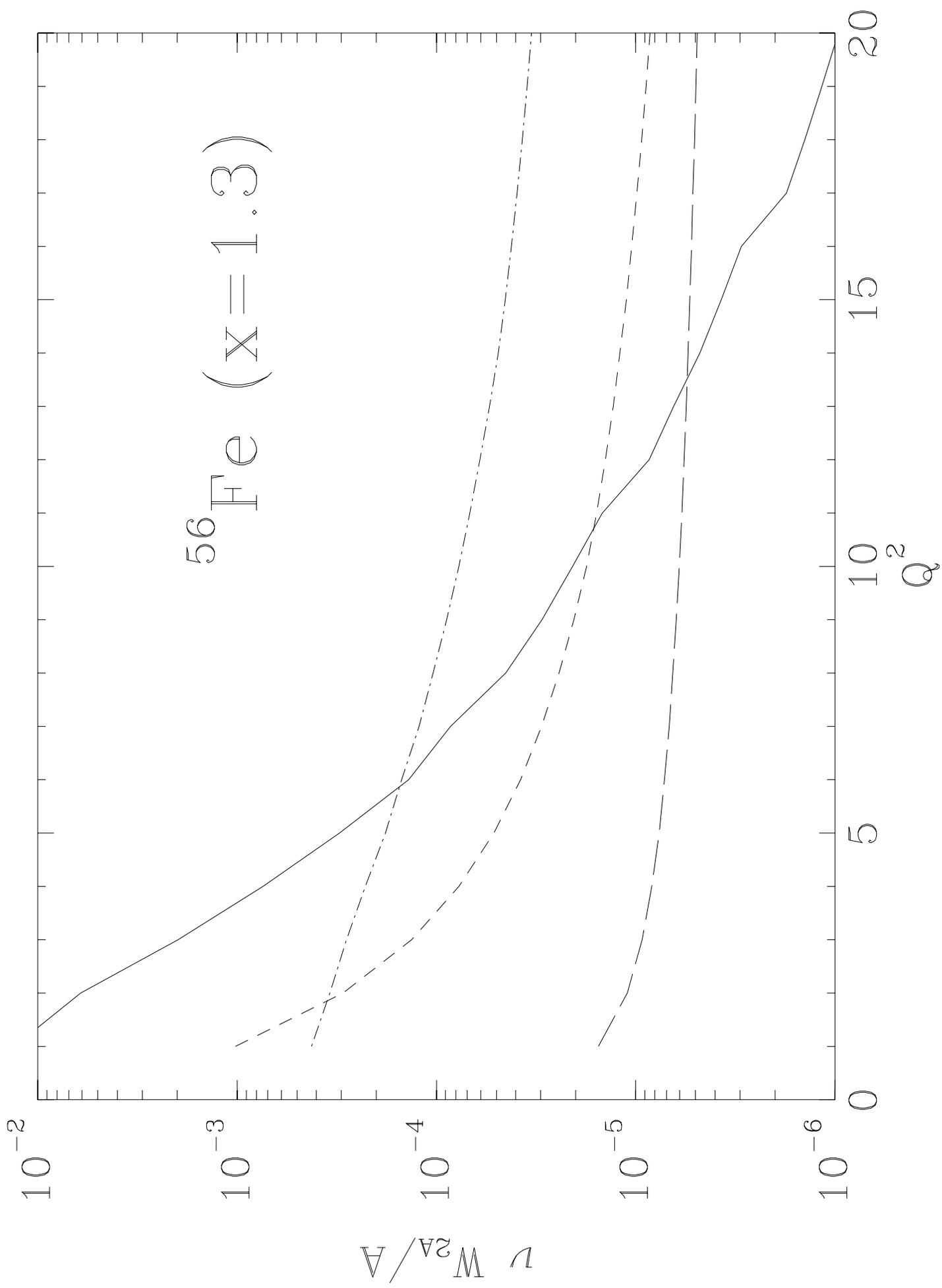
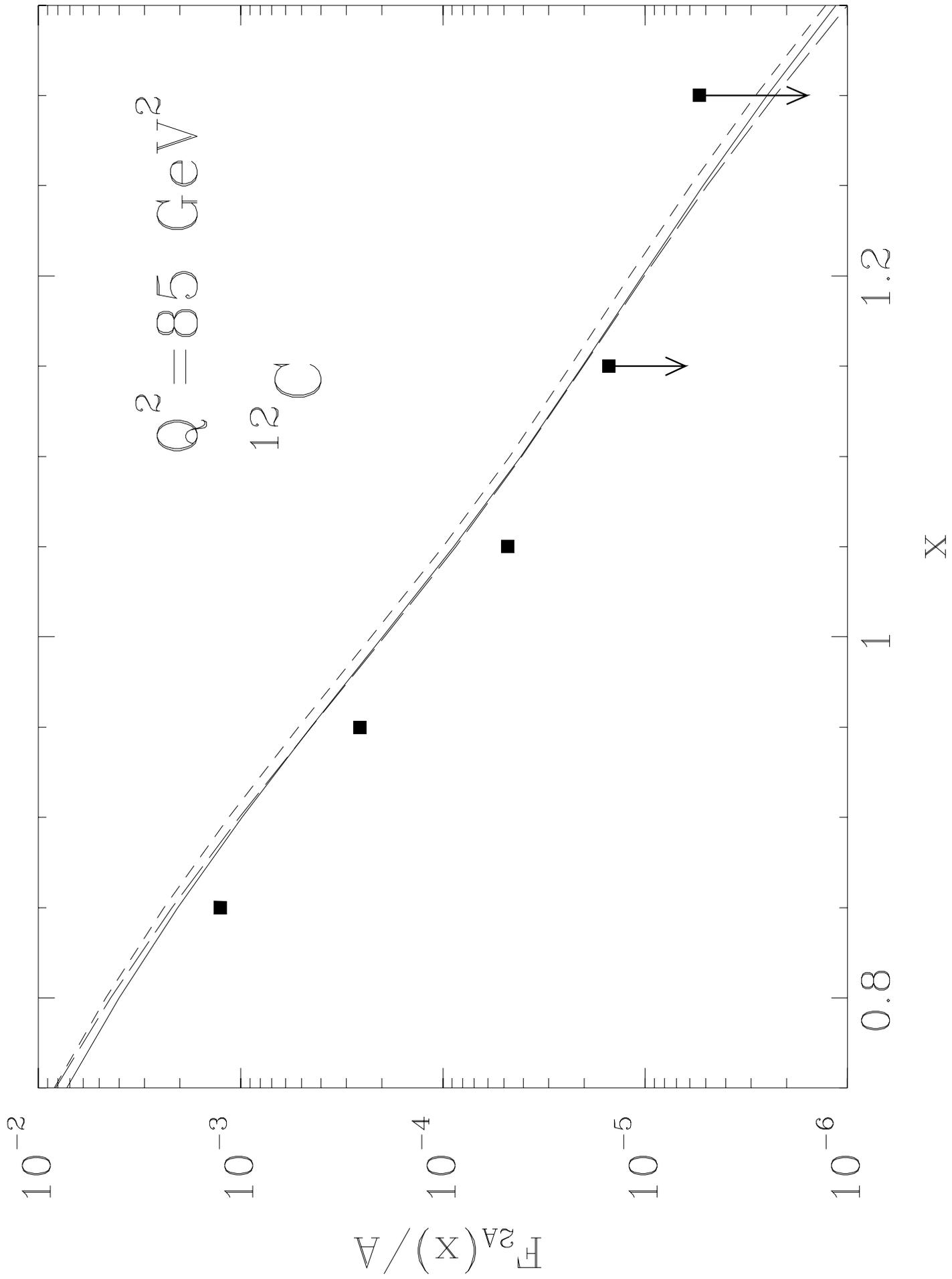
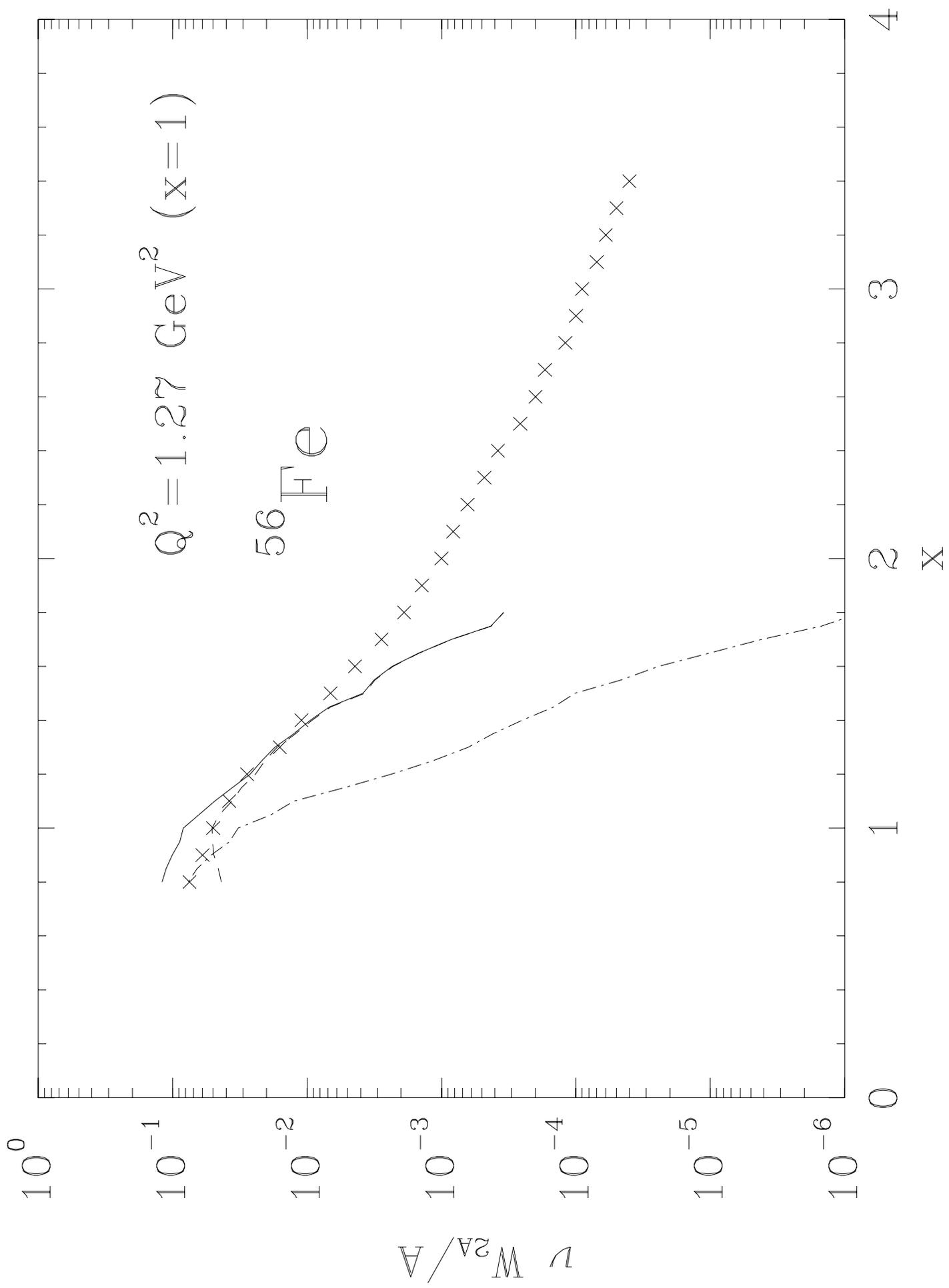
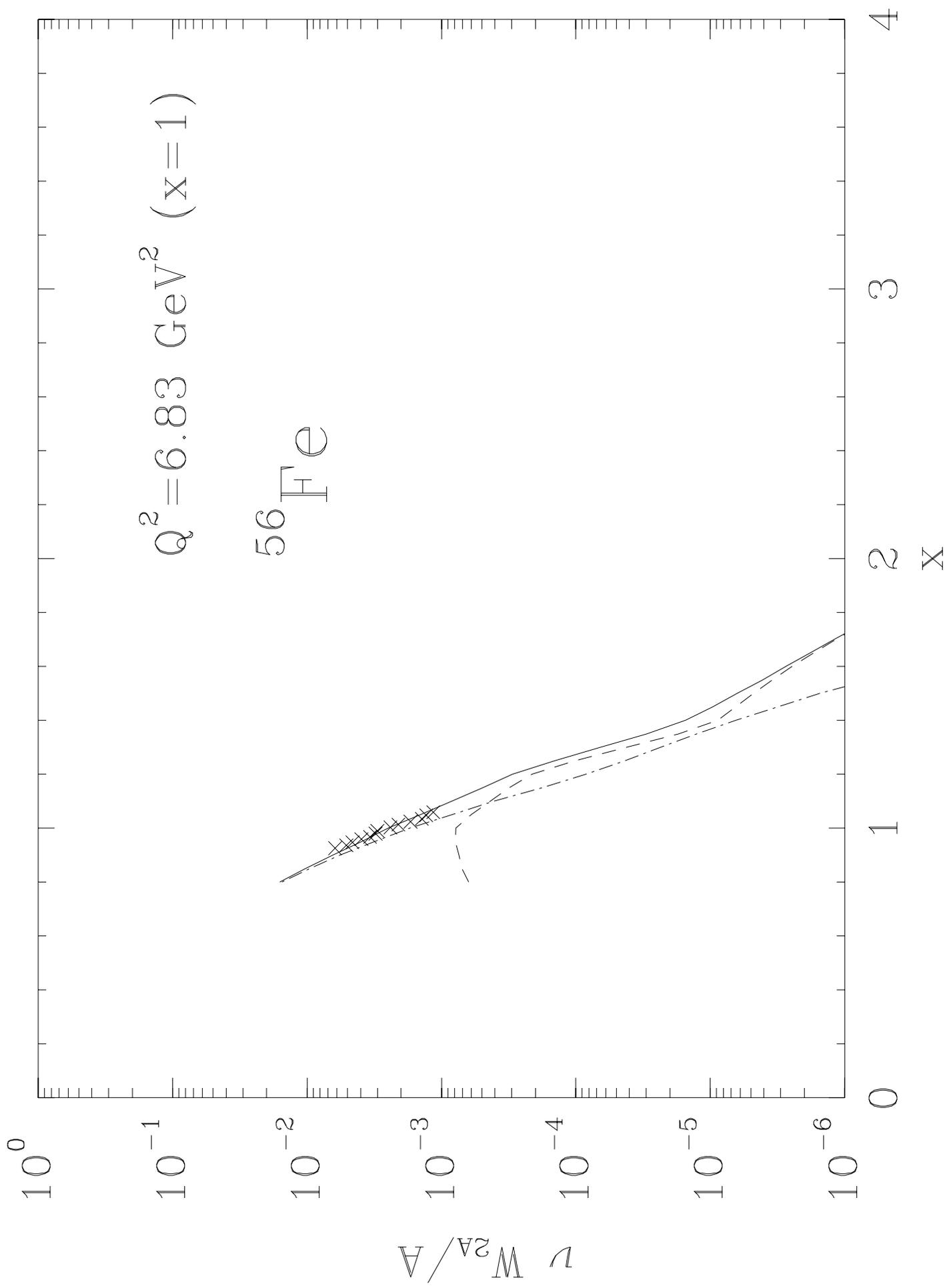
\end{document}